\documentclass[twocolumn]{jpsj3}

\usepackage{txfonts}
\usepackage{graphicx}
\usepackage{dcolumn}
\usepackage{bm}
\usepackage[usenames,dvipsnames]{xcolor}
\usepackage{ulem}
\usepackage{amsmath}
\usepackage{braket}
\usepackage{enumerate}
\usepackage{arydshln}
\usepackage{multirow}
\usepackage{physics}

\usepackage[whole]{bxcjkjatype}

\title{Finite-temperature stability of skyrmion crystals in frustrated magnets: \\
  Role of sixfold anisotropy and uniform spin mode in momentum space}

\author{Kazuki Okigami$^1$ and Satoru Hayami$^2$}
\inst{$^1$Department of Applied Physics, The University of Tokyo, Bunkyo, Tokyo 113-8656, Japan \\
$^2$Graduate school of Science, Hokkaido University, Sapporo 060-0810, Japan}
\abst{We study the finite-temperature stability of skyrmion crystals in frustrated magnets by analyzing the momentum-space exchange interaction of a classical Heisenberg model on a triangular lattice.
Our analysis identifies two key momentum-space features that play a crucial role in stabilizing the skyrmion crystal phase.
The first is the sixfold anisotropy in the momentum-space exchange interaction, which acts as a locking potential favoring triple-$Q$ skyrmion crystals.
Monte Carlo simulations reveal that a larger anisotropy tends to enhance the stability region of the skyrmion crystal in the temperature--magnetic-field phase diagram.
The second factor is the momentum-space energy related to the uniform spin mode, which correlates with the emergence of the skyrmion crystal phase at finite temperatures.
These results provide a further understanding of the stabilization mechanism of the skyrmion crystal phase in frustrated magnets and will be useful for the design of skyrmion-hosting materials.}

\begin{document}
\maketitle

\section{Introduction}
Magnetic skyrmions have attracted significant interest as promising candidates for next-generation devices due to their nontrivial topological features and nanoscale size~\cite{nagaosa2013topological, fert2017magnetic, zhang2020skyrmion}.
These features give rise to emergent phenomena, such as the topological Hall effect~\cite{Ohgushi_PhysRevB.62.R6065, Neubauer_PhysRevLett.102.186602, Hamamoto_PhysRevB.92.115417, kurumaji2019skyrmion} and the topological Nernst effect~\cite{Shiomi_PhysRevB.88.064409, Hirschberger_Nernst_prl2020}.
Moreover, their particle-like nature allows for high controllability at the nanoscale, which is advantageous for device applications such as racetrack memories~\cite{fert2013skyrmions, fert2017magnetic, zhang2020skyrmion}, logic devices~\cite{zhang2015magnetic, Luo2018reconfigurable, Chauwin2019computation}, neuromorphic computing systems~\cite{Kuzum2013synaptic, Huang2017synaptic, Yokouchi2022neuromorphic}, and quantum computing~\cite{Psaroudaki2021SkyrmionQubit, Xia2023qubit}.
To realize these applications, previous research has revealed various methods to manipulate skyrmions, such as electric currents~\cite{Jonietz2010skyrmion, Yu2012skyrmion} and heat gradients~\cite{Yu2021dynamics}, as well as techniques for their creation and annihilation~\cite{Romming2013writing}.

Besides these controls of skyrmions, understanding the stabilization mechanism of skyrmions is also crucial for practical applications.
Although skyrmion crystals (SkX), periodic arrays of skyrmions, were first experimentally observed in noncentrosymmetric materials, such as chiral magnets~\cite{Muhlbauer_2009skyrmion, Neubauer_PhysRevLett.102.186602, yu2010real, adams2010skyrmion, seki2012observation} and polar magnets~\cite{kezsmarki_neel-type_2015, Kurumaji_PhysRevLett.119.237201}, they have also been found in centrosymmetric materials~\cite{kurumaji2019skyrmion, khanh2020nanometric, khanh2022zoology}.
In parallel with these experimental observations, theoretical investigations have revealed that SkX phases can be stabilized in centrosymmetric systems through various mechanisms~\cite{HayamiYambe2024_stabilization}, including magnetic frustration~\cite{Okubo_PhysRevLett.108.017206, leonov2015multiply, Lin_PhysRevB.93.064430, Hayami_PhysRevB.103.224418, kawamura2025frustration}, long-range interactions induced by itinerant electrons~\cite{Martin_PhysRevLett.101.156402,heinze2011spontaneous,takagi2018multiple, Ozawa_PhysRevLett.118.147205,Hayami_PhysRevB.95.224424, Nikolic_PhysRevB.103.155151}, and higher-order interactions~\cite{Hayami_PhysRevB.95.224424, Paul2020role, hayami2021phase}.

While previous theoretical studies have mainly focused on the ground-state properties of the models, understanding the stability of the SkX at finite temperatures is also essential for practical applications in real materials.
For instance, observations of metastable SkXs, which can be long-lived at finite temperatures even out of equilibrium due to topological protection, have been reported in various materials~\cite{Oike2016metastable, karube2016robust, Karube2020metastable}.
Furthermore, in frustrated centrosymmetric models, the ground state is often a single-$Q$ spiral state, and the SkX appears only at finite temperatures via the order-by-disorder mechanism~\cite{Okubo_PhysRevLett.108.017206, mitsumoto2022rkky}.
Therefore, understanding the finite-temperature stability relying on thermal fluctuations is a key issue to be clarified for future applications of skyrmions.

In this paper, we investigate key factors that control the finite-temperature stability of the SkX in frustrated centrosymmetric magnets.
By performing Monte Carlo simulations for the frustrated $J_1$-$J_2$-$J_3$ Heisenberg model on a triangular lattice, which is a minimal model to host the spiral phase and the SkX phase, we find that two parameters derived from the momentum-space exchange interaction are closely related to the finite-temperature SkX stability.
One is the sixfold anisotropy of the exchange interaction in the momentum space dictated by the triangular-lattice symmetry, which acts as a locking potential to stabilize the SkX.
The other is the uniform channel in the momentum-space interaction, which also correlates with the emergence of the SkX phase.
These results shed light on the microscopic mechanism that stabilizes SkXs at finite temperatures, offering guiding principles for designing frustrated magnets hosting SkXs.

The rest of this paper is organized as follows:
In Sec.~II, we present the ground-state analysis of the $J_1$-$J_2$-$J_3$ Heisenberg model on a triangular lattice based on the momentum-space exchange interaction.
In Sec.~III, we first show several Monte Carlo phase diagrams obtained for different sets of exchange interactions in the plane of the temperature and magnetic field.
Then, we examine the influence of sixfold exchange anisotropy on the finite-temperature stability of the SkX phase.
In Sec.~IV, we further investigate the role of the uniform spin mode in the momentum space for the emergence of the SkX.
Finally, we summarize our findings and discuss their implications in Sec.~V.

\section{Model}

\begin{figure}[bt!]
  \begin{center}
    \includegraphics[width=0.8\hsize]{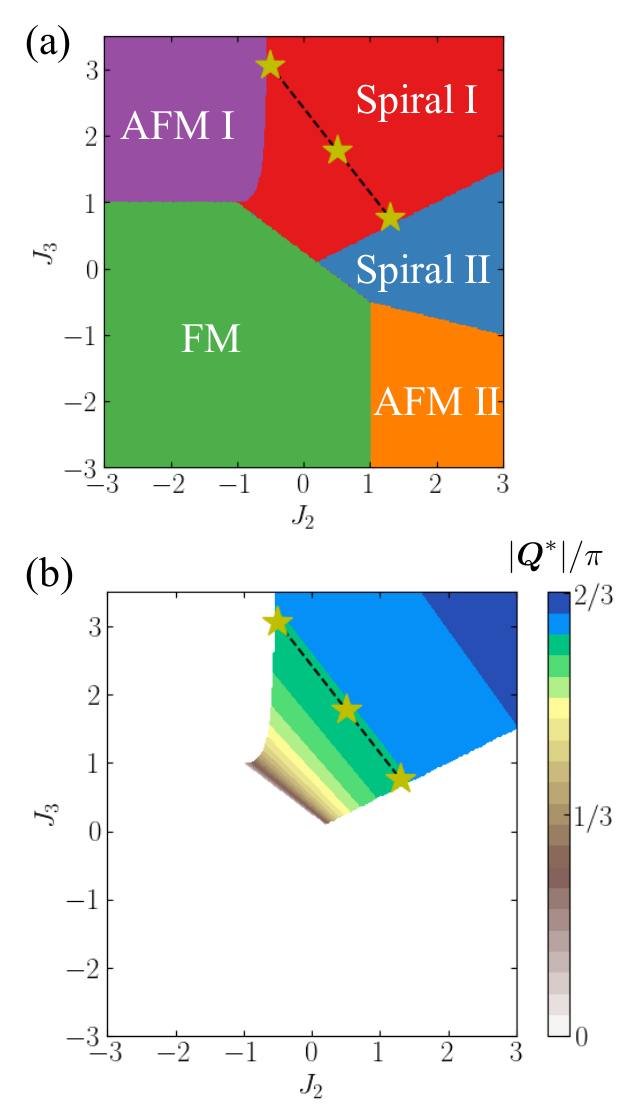}
    \caption{
      \label{fig:PhaseDiagram J1J2J3}
      (a) Ground-state phase diagram of the $J_1$-$J_2$-$J_3$ Heisenberg model on a triangular lattice in the $(J_2, J_3)$ plane with $J_1=-1$.
      (b) Color map of the magnitude of the ordering vector $Q^*=|\bm{Q}^*|$ in the Spiral I phase.
    }
  \end{center}
\end{figure}

We consider a classical $J_1$-$J_2$-$J_3$ Heisenberg model on a two-dimensional triangular lattice.
The Hamiltonian is given by
\begin{equation}
  \mathcal{H} = \sum_{k=1}^{3} \sum_{\langle i,j \rangle_k}
  J_{k} \bm{S}_i \cdot \bm{S}_j
  -H \sum_i S_i^z
  ,\label{eq:J1J2J3 Hamiltonian}
\end{equation}
where $\bm{S}_i$ is a classical spin vector of unit length $|\bm{S}_i|=1$ at site $i$, $J_{k}$ is the $k$-th nearest-neighbor coupling constant, $\langle i,j \rangle_k$ denotes the $k$-th nearest-neighbor pairs, and $H$ is an external magnetic field applied along the $z$-axis.
Throughout this paper, we fix the nearest-neighbor interaction to be ferromagnetic, $J_1 = -1$, which serves as the energy unit, and explore the model's properties by varying $J_2$ and $J_3$.

To analyze the magnetic ground state, we introduce the Fourier transform of the spin variables, $\bm{S}_{\bm{q}} = (1 / \sqrt{N}) \sum_i \bm{S}_i e^{-i \bm{q} \cdot \bm{r}_i}$, where $N$ is the total number of sites, $\bm{r}_i$ is the position vector of site $i$, and $\bm{q}$ is the wave vector.
The exchange part of the Hamiltonian (for $H=0$) can then be expressed in momentum space as
\begin{equation}
  \mathcal{H}_{\rm ex} = \frac{1}{2} \sum_{\bm{q}} J(\bm{q}) \bm{S}_{\bm{q}} \cdot \bm{S}_{-\bm{q}}.
  \label{eq:hamiltonian_q}
\end{equation}
Here, $J(\bm{q})$ is the exchange interaction in the momentum-space.
For the $J_1$-$J_2$-$J_3$ model on the triangular lattice, $J(\bm{q})$ is given by
\begin{align}
  J(\bm{q}) = & \; 2J_1 \left[ \cos q_x + 2\cos\left(\frac{q_x}{2}\right)\cos\left(\frac{\sqrt{3}q_y}{2}\right) \right] \nonumber           \\
  +           & \; 2J_2 \left[ \cos(\sqrt{3}q_y) + 2\cos\left(\frac{3q_x}{2}\right)\cos\left(\frac{\sqrt{3}q_y}{2}\right) \right] \nonumber \\
  +           & \; 2J_3 \left[ \cos(2q_x) + 2\cos(q_x)\cos(\sqrt{3}q_y) \right],
  \label{eq:jq_explicit}
\end{align}
with the lattice constant set to unity.

The ground state in the absence of a magnetic field is determined by the wave vector(s) $\bm{Q}^*$ that minimize $J(\bm{q})$.
This can be shown using the Luttinger-Tisza method~\cite{Luttinger_PhysRev.70.954}, which replaces
the local constraint $|\bm{S}_i|=1$ by a global one $\sum_i |\bm{S}_i|^2 = N$.
Under this approximation, the energy is minimized when only the modes $\bm{S}_{\bm{Q}^*}$ corresponding to the minimum of $J(\bm{q})$ have finite amplitudes.
When this minimum arises at an incommensurate wave vector(s) $\bm{Q}^*$ (and its counterpart $-\bm{Q}^*$), the resulting spin configuration is a spiral state, which satisfies the original local constraint.
In this work, we analyze the features of the $J(\bm{q})$ landscape to understand their contribution to the stability of the SkX at finite temperatures.

By performing this minimization of $J(\bm{q})$ for different sets of $J_2$ and $J_3$, we obtain the classical ground state phase diagram, as shown in Fig.~\ref{fig:PhaseDiagram J1J2J3}(a).
The diagram reveals five distinct magnetic phases:
a ferromagnetic (FM) phase with $\bm{Q}^*=(0,0)$, two commensurate antiferromagnetic phases, AFM I with $\bm{Q}^*=(\pi, 0)$ and AFM II with $\bm{Q}^*=\pi(1, 1/\sqrt{3})$, and two incommensurate spiral phases denoted as Spiral I and Spiral II.
The Spiral I phase is characterized by an ordering wave vector along the $\Gamma$-K direction, $\bm{Q}^* \parallel (1, 0)$, while the Spiral II phase has an ordering wave vector along the $\Gamma$-M direction, $\bm{Q}^* \parallel (\cos(\pi/6), \sin(\pi/6))$.
In this study, we focus our investigation on the properties of the Spiral I phase.
It is noted that the SkX as well as other multiple-$Q$ states are not stabilized as the ground state in the $J_1$-$J_2$-$J_3$ Heisenberg model, since multiple-$Q$ superposition does not satisfy the local constraint $|\bm{S}_i|=1$ without introducing higher-harmonic wave-vector contributions.

\begin{figure*}[ht!]
  \begin{center}
    \includegraphics[width=0.99\hsize]{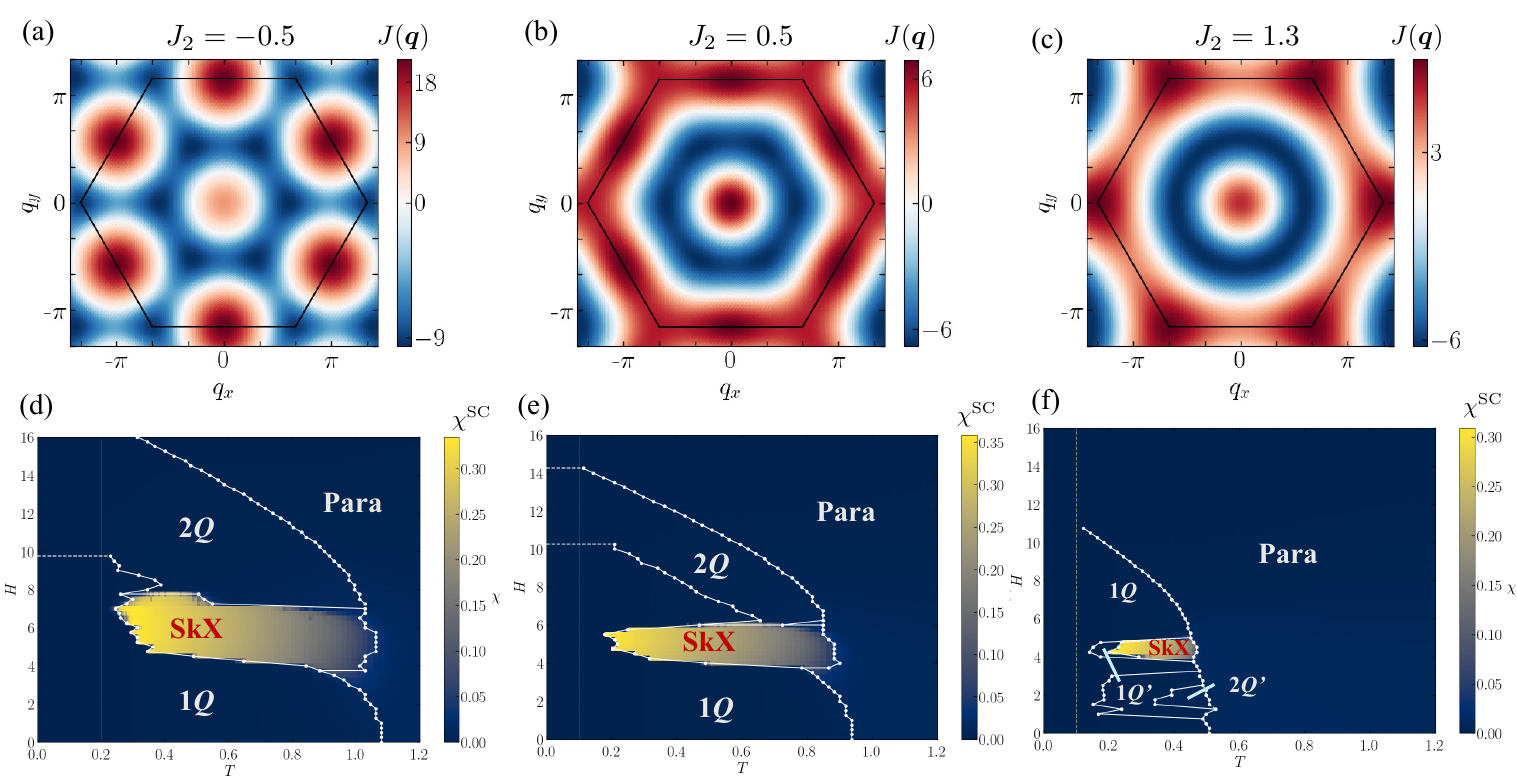}
    \caption{
      \label{fig:Jq PhaseDiagram}
      (a)-(c) Color plots of $J(\bm{q})$ in the momentum space for $Q^*=0.6\pi$ with (a) $J_2=-0.5$ ($J_3=3.061$), (b) $J_2=0.5$ ($J_3=1.7845$), and (c) $J_2=1.3$ ($J_3=0.7638$).
      (d)-(f) Finite-temperature phase diagrams in the plane of temperature $T$ and magnetic field $H$ for the same parameter sets as (a)-(c), respectively.
      Dotted lines indicate minimum temperatures in simulations.
    }
  \end{center}
\end{figure*}

Figure~\ref{fig:PhaseDiagram J1J2J3}(b) presents a color map of the magnitude of the ordering vector, $Q^*=|\bm{Q}^*|$, within the Spiral I region of the phase diagram.
For any given pair of $(J_2, J_3)$ in this phase, $Q^*$ is determined analytically by the condition $\partial J(\bm{q}) / \partial \bm{q} |_{\bm{q}=(Q^*,0)} = 0$.
Then, $Q^*$ satisfies the following relation
\begin{align}
  Q^*    & = \arccos
  \left[
    -\frac{1}{8 J_3}
    \left\{
    (3 J_2 - 2 J_3)
    - \alpha
    \right\}
  \right],                                         \\
  \alpha & = \sqrt{(3 J_2 + 2 J_3)^2 - 8 J_1 J_3}.
\end{align}
For the following analysis of SkX stability at finite temperatures in the next section, we specifically examine a set of parameters for which the ordering vector magnitude is fixed at $Q^*=0.6\pi$.
These parameter sets, which lie on a straight line in the phase diagram as indicated by star symbols in Figs.~\ref{fig:PhaseDiagram J1J2J3}(a) and (b), reveal a linear relationship between $J_2$ and $J_3$ for a fixed $Q^*$.

\section{Role of sixfold anisotropy for skyrmion crystal stability}
\subsection{Finite-Temperature Stability of the Skyrmion Crystal}
While the spiral ground state is characterized by the magnitude of the ordering vector $Q^*$, the stability of a SkX at finite temperatures, which is a superposition of multiple spiral waves (a triple-$Q$ state), is sensitive to the details of the energy landscape $J(\bm{q})$.
To examine the dependence of the stability of the SkX phase on the structure of $J(\bm{q})$, we performed parallel tempering Monte Carlo simulations~\cite{hukushima1996exchange} to obtain the finite-temperature phase diagrams for the three different parameter sets.

To explore the effect of $J(\bm{q})$ landscape, we choose three different values of $J_2$: $-0.5$, $0.5$, and $1.3$, which correspond to $J_3=3.061$, $1.7845$, and $0.7638$, respectively.
The $J(\bm{q})$ plots for these parameter sets are shown in Figs.~\ref{fig:Jq PhaseDiagram}(a)-(c).
The plots demonstrate that the minima of $J(\bm{q})$ are sharp and deep for smaller $J_2$ (e.g., $J_2=-0.5$).

The simulations were conducted on a system of size $N = L^2$ with $L=40$ and $80$ under periodic boundary conditions.
Hereafter, we present results for $L=40$, since the qualitative features remain consistent for both sizes.
For each parameter set, the system was thermalized for $10^6$ to $5 \times 10^6$ Monte Carlo sweeps, followed by $5 \times 10^6$ sweeps for measurements.
The phases are determined by peaks of specific heat and spin structure factors $S^{\alpha}_s(\bm{q}) = (1/N) \sum_{i,j} \expval{S^{\alpha}_i S^{\alpha}_j} e^{i \bm{q} \cdot (\bm{r}_i - \bm{r}_j)}$ ($\alpha = x, y, z$).

The resulting phase diagrams regarding temperature and magnetic field are shown in Figs.~\ref{fig:Jq PhaseDiagram}(d)-(f).
The color map represents the scalar spin chirality, $\chi^{\rm SC} = (1/N) \sum_{i,j,k} \bm{S}_i \cdot (\bm{S}_j \times \bm{S}_k)$, which represents the presence of SkX phase.
Here, the summation runs over all triangular plaquettes in the lattice.

In addition to the SkX phase, we identify topologically trivial phases with $\chi^{\rm SC}=0$, including the single-$Q$ spiral (1$Q$) phase and a double-$Q$ spiral (2$Q$) phase, which consists of two in-plane spiral components, one out-of-plane spiral and ferromagnetic component.
Furthermore, for $J_2=1.3$ (Fig.~\ref{fig:Jq PhaseDiagram}(f)), the phase diagram becomes more complicated.
The SkX phase appears only in a very narrow field window.
In this regime, we also observe additional competing phases, labeled 1$Q$' and 2$Q$', where the dominant ordering vectors are slightly shifted from the high-symmetry points.

Among those sets of parameters, we find that the SkX phase is stabilized at finite temperatures in all cases, but its stability region varies significantly depending on the parameter set.
For instance, the SkX phase appears in a wide range of temperature and magnetic field for $J_2=-0.5$ (Fig.~\ref{fig:Jq PhaseDiagram}(d)), while it is confined to a narrow region for $J_2=1.3$ (Fig.~\ref{fig:Jq PhaseDiagram}(f)).
To understand the origin of this variation, we analyze the sixfold anisotropy of $J(\bm{q})$ in the next subsection.

\subsection{Sixfold Anisotropy in Momentum-Space Interaction}

\begin{figure}[bt!]
  \begin{center}
    \includegraphics[width=0.7\hsize]{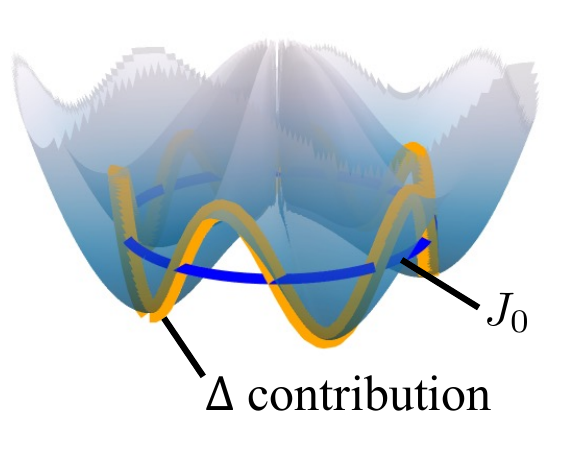}
    \caption{
      \label{fig:Delta schematic}
      Schematic image of the sixfold anisotropy $\Delta$ and the isotropic part $J_0$ of $J(\bm{q})$.
    }
  \end{center}
\end{figure}

Due to the discrete $C_6$ rotational symmetry of the triangular lattice, $J(\bm{q})$ is not isotropic for a fixed $|\bm{q}|=Q^*$.
Indeed, it exhibits a sixfold periodic modulation~\cite{Hayami_PhysRevB.93.184413} when expressed in polar coordinates $\bm{q}=(Q^*, \phi)$.
This anisotropic energy landscape can be characterized by
\begin{equation}
  J(Q^*, \phi) = J_0 - \Delta \cos(6\phi),
  \label{eq:Jq_anisotropy}
\end{equation}
where $J_0$ is the isotropic average part of the exchange energy on the ring at $|\bm{q}|=Q^*$, and $\Delta$ is the amplitude of the sixfold anisotropy.
The coefficient $\Delta$ is formally defined as
\begin{equation}
  \Delta = \frac{1}{\pi} \int_{-\pi}^{\pi} J(Q^*, \phi) \cos(6\phi) d\phi.
  \label{eq:Delta_definition}
\end{equation}

This anisotropy is schematically illustrated in Fig.~\ref{fig:Delta schematic}.
The energy $J(\bm{q})$ along the circle $|\bm{q}|=Q^*$ has six minima at angles corresponding to the high-symmetry directions of the lattice (e.g., $\Gamma$-K directions).
A positive $\Delta$ ensures that these directions are indeed the energy minima.
The magnitude $|\Delta|$ represents the energy depth of these minima compared to the maxima.

The presence of this anisotropy is crucial for the formation of a triple-$Q$ state, such as the SkX.
The term $\Delta \cos(6\phi)$ acts as a locking potential that energetically favors the selection of three ordering vectors $\bm{Q}_1, \bm{Q}_2$, and $\bm{Q}_3$, which are connected by the $C_6$ symmetry and correspond to the minima of $J(\bm{q})$.
The larger the magnitude of this locking potential $\Delta$ is, the stronger the energetic gain for the system to form a stable triple-$Q$ structure, thereby enhancing the stability of the SkX.

\begin{figure}[bt!]
  \begin{center}
    \includegraphics[width=0.99\hsize]{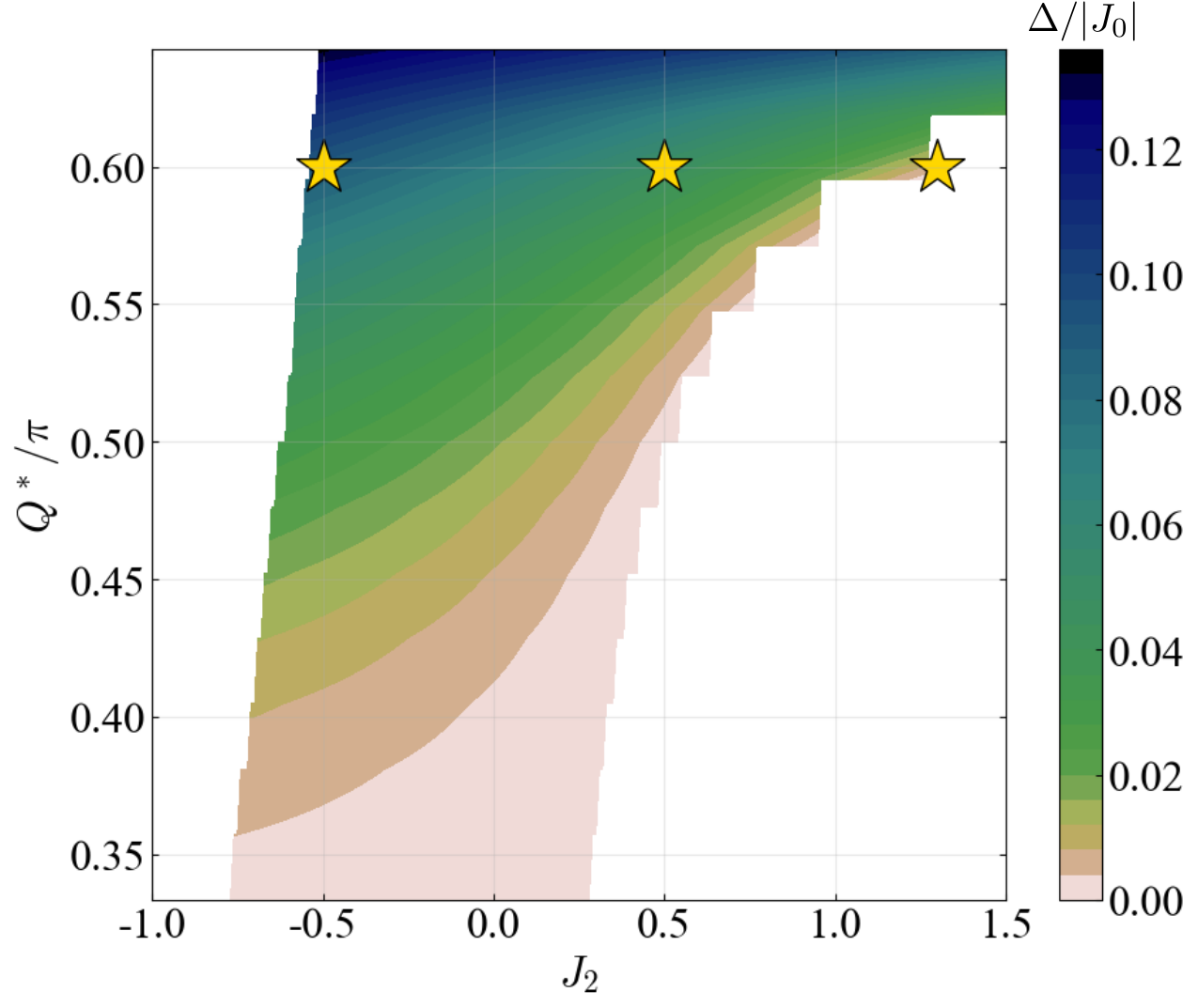}
    \caption{
      \label{fig:Delta J2 Q dependence}
      $J_2$ and $Q^*$ dependence of the sixfold anisotropy $\Delta$ normalized by $|J_0|$.
      The star symbols indicate the parameter sets used in Fig.~\ref{fig:Jq PhaseDiagram}, which correspond to $Q^*=0.6\pi$ with $J_2=-0.5, 0.5,$ and $1.3$ that gives $\Delta=1.440$, $0.672$, and $0.050$, respectively.
    }
  \end{center}
\end{figure}

To investigate the role of sixfold anisotropy, which we suppose is crucial for SkX stability, we calculated its amplitude $\Delta$ as a function of the model parameters.
As previously mentioned, for any pair of $(J_2, Q^*)$ within the Spiral I phase, $J_3$ is uniquely determined.
This allows us to map the behavior of $\Delta$ as functions of $J_2$ and $Q^*$.

Figure~\ref{fig:Delta J2 Q dependence} shows a color plot of the normalized sixfold anisotropy, $\Delta/|J_0|$, in the $(J_2, Q^*)$ plane.
Firstly, for a fixed $Q^*$, the magnitude of the anisotropy $\Delta$ decreases as $J_2$ increases.
This reduction in the locking potential suggests a corresponding decrease in the stability of the SkX for larger $J_2$.
Secondly, $\Delta$ also tends to be weaker for smaller ordering vector magnitudes, $Q^*$.

As we have previously discussed, the difference in the momentum-space sixfold anisotropy $\Delta$ for different $J_2$ values directly influences $J(\bm{q})$ landscape and the stability of the SkX phase.
By revisiting $J(\bm{q})$ plots in Figs.~\ref{fig:Jq PhaseDiagram}(a)-(c), we can observe the decreasing tendency of $\Delta$ directly.
While $J(\bm{Q}^*)$ exhibits sharp minima for smaller $J_2$, the landscape around the minima becomes flatter and more rounded as $J_2$ increases.
This confirms that the sixfold anisotropy, and thus the energetic preference for the SkX state, is significantly suppressed by increasing the next-nearest-neighbor interaction $J_2$.
Projecting this understanding onto the phase diagrams in Figs.~\ref{fig:Jq PhaseDiagram}(d)-(f), we find that the SkX phase is most stable for $J_2=-0.5$, where $\Delta$ is largest, and its stability region shrinks as $J_2$ increases, consistent with the decreasing $\Delta$.

\section{Role of the Uniform Spin Mode}

\begin{figure}[bt!]
  \begin{center}
    \includegraphics[width=0.99\hsize]{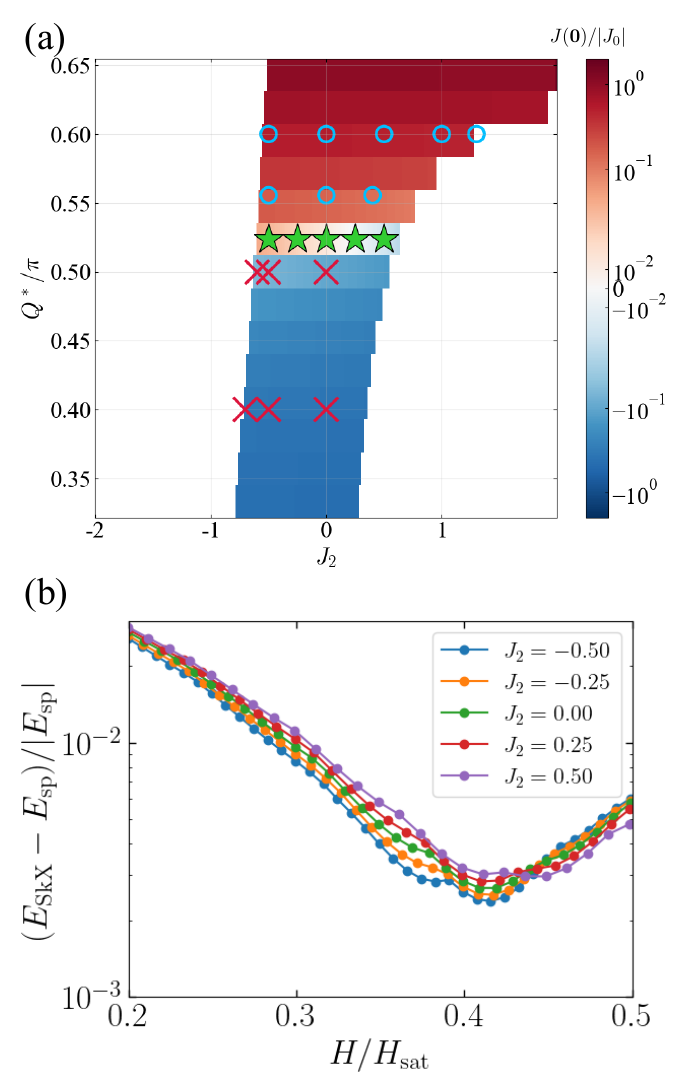}
    \caption{
      \label{fig:J0 J2 Q dependence}
      (a) Color plot of $J(\bm{0})/|J_0|$ in the $(J_2, Q^*)$ plane. Circles and crosses represent the presence and absence of the SkX phase at finite temperatures, respectively.
      The star symbols indicate the parameter sets used in Fig.~\ref{fig:J0 T dependence}, which correspond to $Q^*=11\pi/21$ with $J_2=-0.5, -0.25, 0, 0.25,$ and $0.5$.
      (b) Relative energy difference between the SkX and spiral states as a function of magnetic field $H$ for $Q^*=11\pi/21$ with various $J_2$.
    }
  \end{center}
\end{figure}

\begin{figure}[hbt!]
  \begin{center}
    \includegraphics[width=0.9\hsize]{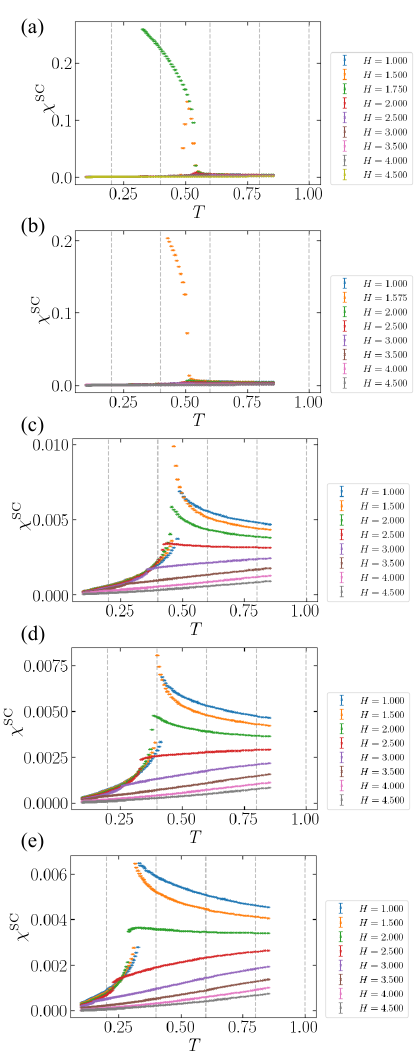}
    \caption{
      \label{fig:J0 T dependence}
      (a)--(e) Temperature dependence of the scalar spin chirality $\chi^{\rm SC}$ at various magnetic fields for $J_2=-0.5, -0.25, 0, 0.25,$ and $0.5$, respectively.
    }
  \end{center}
\end{figure}

\begin{figure}[hbt!]
  \begin{center}
    \includegraphics[width=0.99\hsize]{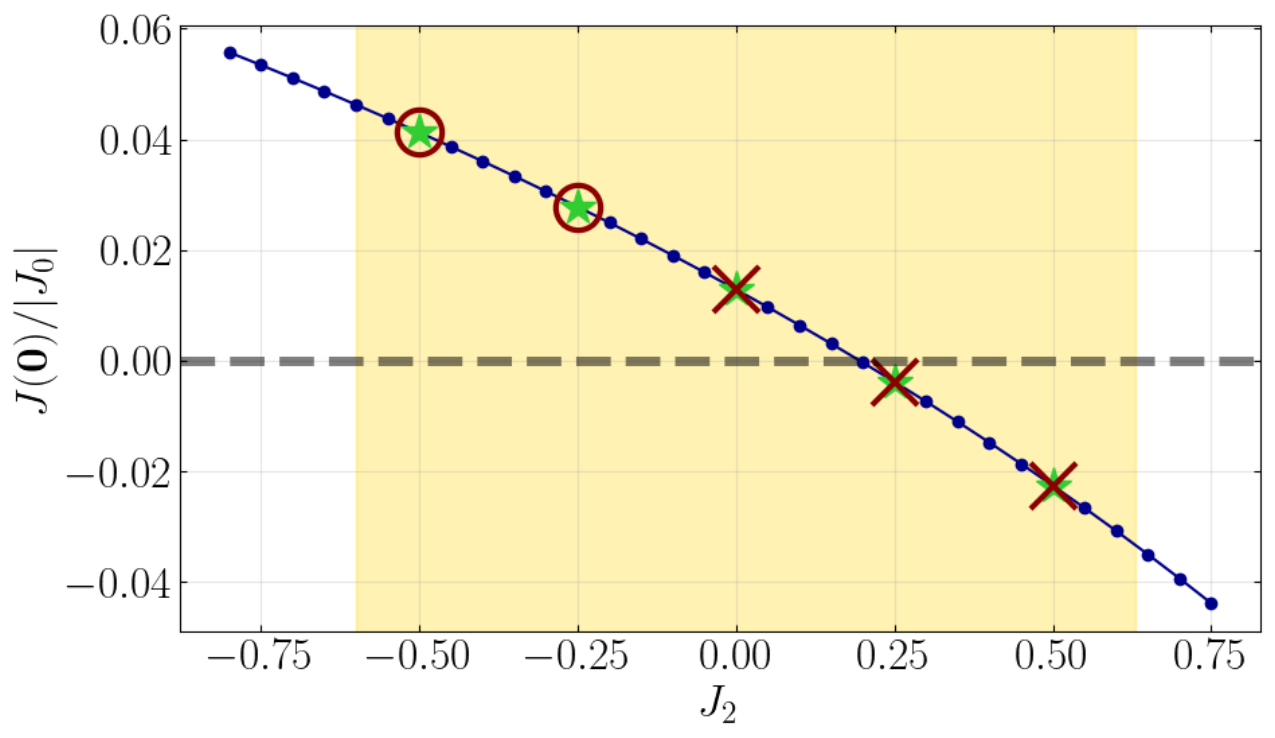}
    \caption{
      \label{fig:J0 J2 dependence}
      $J_2$ dependence of $J(\bm{0})/|J_0|$ for $Q^*=11\pi/21$.
      The star symbols indicate the parameter sets used in Fig.~\ref{fig:J0 T dependence}.
      Circles and crosses represent the presence and absence of the SkX phase at finite temperatures, respectively.
      The region drawn in yellow stands for the Spiral I phase for $Q^*=11\pi/21$.
    }
  \end{center}
\end{figure}

In addition to the sixfold anisotropy which enhances the SkX state stability, we find another key feature of the energy landscape: the energy of the uniform spin mode, $J(\bm{q}=\bm{0})$.
This term, hereafter denoted as $J(\bm{0})$, governs the energy associated with a uniform component and spontaneous magnetization in the spin structure.

To understand its behavior across the parameter space, we calculated $J(\bm{0})$ as functions of $Q^*$ and $J_2$.
Figure~\ref{fig:J0 J2 Q dependence}(a) shows a color plot of the normalized value, $J(\bm{0})/|J_0|$.
The plot reveals that $J(\bm{0})$ becomes large and positive for large $Q^*$, while it takes negative values for small $Q^*$.
A weaker dependence is also seen for $J_2$, where larger (more positive) $J_2$ values tend to decrease $J(\bm{0})$.

Importantly, Fig.~\ref{fig:J0 J2 Q dependence}(a) is presented with markers indicating the emergence (circles) or absence (crosses) of a finite-temperature SkX phase, as determined by our simulations.
A clear correlation emerges: the SkX phase tends to appear when $J(\bm{0})$ is larger and is suppressed when $J(\bm{0})$ becomes smaller.
To examine this relationship, we focus on a specific parameter line where $J(\bm{0})$ changes its sign as $J_2$ is varied, choosing a fixed ordering vector of $Q^* = 11\pi/21$.

First, we performed the variational analysis comparing the energy of the SkX state ($E_{\rm SkX}$) and the spiral state ($E_{\rm sp}$).
Figure~\ref{fig:J0 J2 Q dependence}(b) plots the relative energy difference, $(E_{\rm SkX}-E_{\rm sp}) / |E_{\rm sp}|$, as a function of $H$ normalized by the saturation field $H_{\rm sat} = -(J(\bm{Q}^*) - J(\bm{0}))$.
The result shows that the energy difference is smaller for smaller $J_2$ (where $J(\bm{0})$ is larger) around $H / H_{\rm sat} \simeq 0.3$, where the SkX phase is likely to appear, indicating that the SkX state is energetically more competitive in this regime.
Nevertheless, the spiral state remains the ground state ($E_{\rm sp} < E_{\rm SkX}$) for all parameters considered.

To investigate the finite-temperature behavior, we performed Monte Carlo simulations for five parameter sets along the line $J_2=-0.5, -0.25, 0, 0.25, 0.5$, and $J_3=1.5713, 1.296,  1.02, 0.7443, 0.4687$ on a system with $L=84$, indicated by star symbols in Figs.~\ref{fig:J0 J2 Q dependence}(a) and \ref{fig:J0 J2 dependence}.
The results are summarized in Fig.~\ref{fig:J0 T dependence}.
Figures~\ref{fig:J0 T dependence}(a)--(e) show the temperature $T$ dependence of the scalar spin chirality $\chi^{\rm SC}$ for parameter sets above, respectively, at various magnetic fields.
A finite $\chi^{\rm SC}$ is clearly observed for $J_2=-0.5$ and $-0.25$, confirming the appearance of the SkX phase.
In contrast, for $J_2 \ge 0$, $\chi^{\rm SC}$ remains almost zero, indicating the absence of the SkX.
This result demonstrates a transition from the presence to the absence of the SkX phase by changing $J_2$.

This transition is directly correlated with the value of $J(\bm{0})$.
Figure~\ref{fig:J0 J2 dependence} plots $J(\bm{0})/|J_0|$ as a function of $J_2$ for our chosen $Q^*$.
Circles and crosses again indicate the presence and absence of the SkX phase, respectively.
The SkX phase disappears when $J(\bm{0})/|J_0|$ falls below a small positive threshold $J(\bm{0})/|J_0| \sim 0.02$.
This finding suggests that $J(\bm{0})$ can act as a crucial parameter governing the emergence of the SkX phase at finite temperatures in frustrated Heisenberg models.

\section{Summary and Discussion}
We have explored key factors that influence the finite-temperature stability of the SkX in the classical $J_1$-$J_2$-$J_3$ Heisenberg model on a triangular lattice.
We started by introducing the ground-state analysis of the model based on the momentum-space exchange interaction $J(\bm{q})$.
By performing Monte Carlo simulations for several parameter sets with different $J(\bm{q})$ landscapes, we obtained finite-temperature phase diagrams and examined the stability of the SkX phase.
We first identified the momentum-space sixfold anisotropy $\Delta$ of the model induced by the triangular lattice geometry acts as a locking potential to stabilize the SkX state for broader stability region with regard to temperature and magnetic field.
Secondly, we found that the energy of the uniform spin mode $J(\bm{0})$ also plays a key role in the emergence of the SkX phase by comparing several parameter sets where $J(\bm{0})$ varies its value.

It is also important to note that these two factors, $J(\bm{0})$ and $\Delta$, are likely correlated.
Indeed, both $\Delta$ and $J(\bm{0})$ becomes larger when $J_2$ is small as we have seen above.
Therefore, by combining these insights, we can understand that the SkX phase is favored since the entropic stabilization by thermal fluctuation and comparative energetic gain against the single-$Q$ state are complementarily enhanced.

We propose that these two parameters can benefit the SkX stability.
This framework is not limited to the frustrated Heisenberg model but can be applied to any system, including those with the DM interaction or the long-range RKKY interaction, by analyzing their respective momentum-space energy landscapes.
These findings provide a systematic understanding of SkX stability in frustrated magnets and offer guiding principles for discovering suitable materials hosting robust skyrmion phases.

\section*{Acknowledgments}
This work was supported by JST SPRING, Grant Number JPMJSP2108.
This was also supported by JSPS KAKENHI Grants Numbers JP21H01037, JP22H04468, JP22H00101, JP22H01183, JP23H04869, JP23K03288, JP23K20827, and by JST PRESTO (JPMJPR20L8) and JST CREST (JPMJCR23O4).

\bibliographystyle{jpsj}
\bibliography{ref}

\end{document}